\documentclass[12pt]{iopart} 

\usepackage{iopams}
\usepackage{setstack}
\usepackage{graphicx}
\usepackage{amsfonts}

\newcommand{\BEA}{\begin{eqnarray}}
\newcommand{\EEA}{\end{eqnarray}}

\usepackage{mathrsfs}
\usepackage{color}

\newcommand{\abs}[1]{\left| #1\right|} 
\newcommand{\ket}[1]{\vert#1\rangle} 
\newcommand{\bra}[1]{\langle#1\vert} 
\newcommand{\brak}[2]{\langle#1\vert#2\rangle}

\newcommand{\hk}{\hat{k}}

\newcommand{\hx}{\hat{x}}
\newcommand{\hy}{\hat{y}}

\newcommand{\hsfA}{\,\,\hat{\boldsymbol{\!\!\mathcal{A}}}}
\newcommand{\sfA}{\boldsymbol{\mathcal{A}}}
\newcommand{\bsfA}{\,\,\bar{\boldsymbol{\!\!\mathcal{A}}}}

\newcommand{\bE}{\boldsymbol{E}}
\newcommand{\bK}{\boldsymbol{K}}
\newcommand{\bR}{\boldsymbol{R}}
\newcommand{\bV}{\boldsymbol{V}}

\newcommand{\be}{\boldsymbol{e}}
\newcommand{\bk}{\boldsymbol{k}}
\newcommand{\bor}{\boldsymbol{r}}

\newcommand{\obR}{\bar{\bR}}
\newcommand{\opsi}{\bar{\psi}}

\newcommand{\bone}{\boldsymbol{1}}
\newcommand{\bhone}{\hat{\boldsymbol{1}}}

\newcommand{\hbR}{\hat{\boldsymbol{R}}}
\newcommand{\hbK}{\hat{\boldsymbol{K}}}

\newcommand{\wpsi}{\widetilde{\psi}}

\begin{document}

\title{Goos-H\"anchen and Imbert-Fedorov shifts from a quantum-mechanical perspective}
\author{Falk T\"{o}ppel$^{1,2,3}$, Marco Ornigotti$^{1,2}$ and Andrea Aiello$^{1,2}$ 
}
\author{}
\ead{falk.toeppel@mpl.mpg.de} 
\address{$^1$ Max Planck Institute for the Science of Light, G\"{u}nther-Scharowsky-Stra{\ss}e 1/Bldg. 24, 91058 Erlangen, Germany}
\address{$^2$ Institute for Optics, Information and Photonics, Universit\"{a}t Erlangen-N\"{u}rnberg, Staudtstra{\ss}e 7/B2, 91058 Erlangen, Germany}
\address{$^3$ Erlangen Graduate School in Advanced Optical Technologies (SAOT), Paul-Gordan-Straße 6, 91052 Erlangen, Germany}
\date{\today}

\begin{abstract}
We study the classical optics effects known as Goos-H\"anchen and Imbert-Fedorov shifts, occurring when reflecting a bounded light beam from a planar surface, by using a quantum-mechanical formalism. 
This new approach allows us to naturally separate the spatial shift into two parts, one independent on orbital angular momentum (OAM) and the other one showing OAM-induced spatial-vs-angular shift mixing. In addition, within this quantum-mechanical-like formalism, it becomes apparent that the angular shift is proportional to the beams angular spread, namely to the variance of the transverse components of the wave vector. Moreover, we extent our treatment to the enhancement of beam shifts via weak measurements and relate our results to recent experiments. 
\end{abstract}


\maketitle

\section{Introduction}
Theoretical physics is all about describing nature in terms of mathematics. However, often there is more than one way to do so. Therefore, different mathematical formalisms are frequently developed in physics to describe one and the same physical phenomenon. In fact, this is even useful since each description offers its own viewpoint onto the underlying physics and some viewpoints are more suited to observe certain details than others. Famous examples of this fact include Newtonian and Hamiltonian formulation of classical mechanics \cite{goldstein} or the Heisenberg and Schr\"odinger picture in quantum mechanics \cite{sakurai}. While the Newtonian mechanics relies on the system's forces, Hamiltonian mechanics rests on the system's energy. Moreover, while the Heisenberg picture resembles classical dynamics, the Schr\"odinger picture stresses the wave-like properties of quantum particles, e. g., electrons.

Following the idea of changing the perspective, in this work we propose a different treatment of beam shifts, by describing this purely classical phenomenon with the mathematical formalism of proper quantum mechanics (QM). 
Already in 1987 M. A. Player used a QM-like formalism to calculate transverse beam shifts as expectation values of Hermitean operators \cite{Player}. Earlier derivations of beam shifts using a classical treatment can be found in \cite{Schilling, Fedoseev, Liberman}. Quite later, in 2004 Onoda and coworkers \cite{Onoda} used the concept of Berry phase and QM conservation laws to predict the existence of a Hall effect of light. A similar but distinct treatment of this phenomenon was also furnished by Bliokh\&Bliokh \cite{B&B,B&B2007}.
From experimental point of view, the connection between classical beam shifts and quantum mechanics was probably first  exploited in 2008 by Hosten and Kwiat \cite{hosten} who used a well known quantum-weak-measurement technique to measure the spin Hall effect of light occurring in optical refraction. However, shortly afterwards Aiello and Woerdman \cite{aiello2008}  showed that such a connection has a pure formal character and that the Hosten\&Kwiat experiment also admits a  fully classical optics description. 
In 2009  Aiello \emph{et al.}, used again a QM formalism to illustrate the ``duality'' existing between spatial and angular beam shifts \cite{aiello2009} and in 2010 Merano \emph{et al.} \cite{merano} had written spatial and angular real-valued physical shifts as weighted sums of real and imaginary  parts of  complex-valued shifts, respectively. This formalism was further elaborated  in early 2012 by Aiello \cite{njp_gh_if}. Finally, later in 2012 Dennis and G\"otte in two excellent papers \cite{gotte1,gotte2} provided for a unified view of all polarization-dependent beam shift phenomena still by exploiting classical/quantum analogies.

The aim of the present work is to move a step forward in the ``quantum'' direction by adopting an ``ab initio'' quantum formalism for a unified description of \emph{all} beam shift phenomena. By exploiting the formal analogy between the paraxial wave equation and the two-dimensional Schr\"{o}dinger equation \cite{stoler} we can represent beam propagation as 
a  ``time'' evolution generated by displacement operator quadratic in the ``momentum'' operator and, therefore, calculate both spatial and angular shifts by using a common formalism. As it will be shown later, in this manner all beam shift phenomena will manifest a natural connection.

\begin{figure}[h!]
\begin{center}
\includegraphics[angle=0,width=12.0truecm]{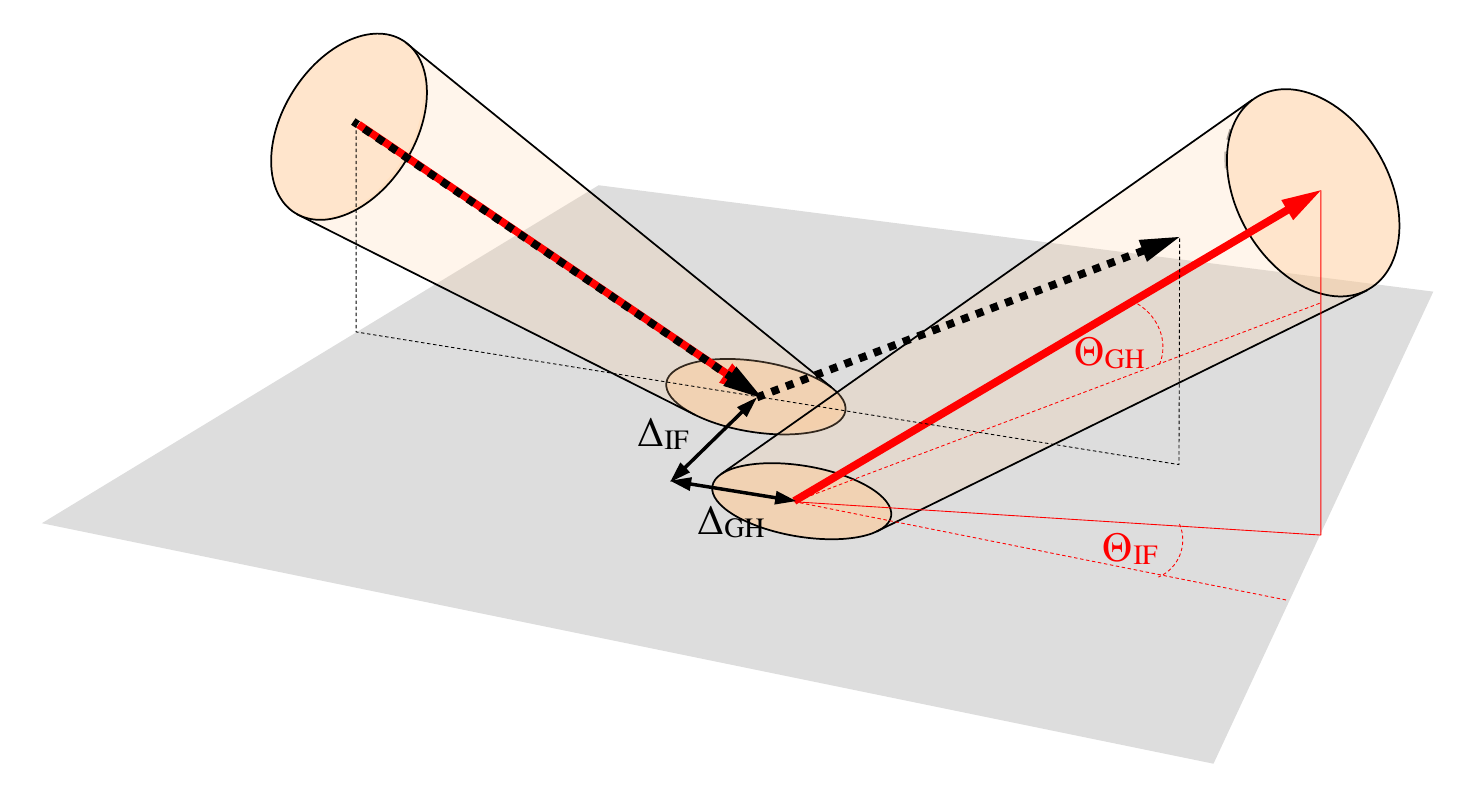}
\caption{\label{fig.shifts} (color online) Visualization of the beam shifts occurring upon reflection of a bounded beam from a planar surface.}
\end{center}
\end{figure}

Beam shifts are deviations from geometric optics (ray optics) predictions that a beam with finite transverse extent experiences on reflection and/or refraction from a planar surface  (see figure \ref{fig.shifts}). The Goos-H\"anchen (GH) \cite{gh_orig,artmann,chiu,guirk} and Imbert-Fedorov (IF) shifts \cite{fedorov_orig,imbert_orig,costa,pillon} are the most celebrated examples thereof (see figure \ref{fig.shifts}). The spatial and angular shifts do occur in the plane of incidence (GH shifts) as well as orthogonal to the plane of incidence (IF shifts). 
%
%
Beam shifts have been subject of extensive studies in the past decades both theoretically and experimentally for different kind of surfaces \cite{ref1a,ref1b,ref1c,ref1d}, different beam shapes \cite{merano,ref2a,ref2b,ref2d} and they have also been recently studied for the non-monochromatic case \cite{marco}. For a detailed review and further information on this topic we refer the reader to \cite{njp_gh_if,shift_overview} and references therein.

One of the main advantages given by a quantum mechanical approach to description of beam shift phenomena, is that it furnishes in a natural manner the reason why the angular GH and IF shifts are proportional to the angular aperture of the beam, i.e. proportional to the variance of the transverse component of the wave vector. Furthermore, in this context also clearly appears that the spatial shifts naturally separate into two terms, one independent of and the other dependent on the beams orbital angular momentum (OAM).

We proceed with the following agenda: Section 2 fixes the notation for the solution of the paraxial wave equation. Hereafter, these results are translated into a quantum-mechanical notation (section 3). The main part of this work is contained in section 4. There, the problem of the beam shifts is studied in terms of the quantum-mechanical formalism derived earlier, including a description of the enhancement of the beam shifts through weak measurements. It follows section 5 where we discuss our results and relate them to experiments on weak measurements. We conclude our paper with some final remarks in section 6.

%
%
%
%
%

%
\section{Paraxial wave equation: classical optics notation}
The scalar wave equation of a monochromatic electromagnetic field${}^1$\footnote[0]{$\!\!\!\!\!\!\!{}^1$\;\;Please note that here we follow the notation introduced in \cite{berry}, i.e. we represent three dimensional vectors by a lowercase boldface symbols, e.g., $\bor=\{x,y,z\}=\{x_1,x_2,x_3\}$, and by  capital boldface symbols its transverse components, e.g., $\bor=\{\bR,z\}$ with $\bR=\{x,y\}=\{x_1,x_2\}$.} $E(\bor,t)=u(x,y,z)\, e ^{ i  k(z - c t)}$ propagating in free space mainly along the $z$ direction is well approximated by the paraxial wave equation \cite{siegman}
\BEA\label{paraxEq1}
\left(\frac{\partial^2}{\partial x^2} + \frac{\partial^2}{\partial y^2} + 2  i  k\frac{\partial}{\partial z}\right)u(x,y,z)=0,
\EEA
with $k=\abs{\bk}>0$ being the modulus of the wave vector $\bk$. It is well known that the normalized fundamental solution $f(x,y,z)$ of this equation is a Gaussian beam:

%
\BEA\label{paraxSolFund}
f(x,y,z) = \sqrt{\frac{k L}{\pi}} \frac{1}{z -  i  L}\exp\left(\frac{ i  k}{2}\frac{x^2 + y^2}{z -  i  L}\right).
\EEA
%
%
%
Here $L>0$ is an arbitrary length (commonly known in optics as the Rayleigh range) that fixes the width of the intensity distribution $\abs{f(x,y,z)}^2$ 
\BEA\label{varianceGauss2}
\langle x^2 \rangle \equiv \int\!\!\!\!\int x^2 \abs{f(x,y,z)}^2 \rmd x  \rmd y = \frac{z^2 + L^2}{2 k L}=\langle y^2 \rangle,
\EEA
evaluated at $z=0$:
\BEA\label{varianceGaussZero}
\left. \langle x^2 \rangle \right|_{z=0} = \left. \langle y^2 \rangle \right|_{z=0} = \frac{ L}{2 k } \equiv \frac{w_0^2}{4},
\EEA
where we have introduced the so-called \emph{waist} of the beam $w_0>0$. 
%
%

It is also useful to introduce the two-dimensional Fourier representation, i.e. the angular spectrum, of the fundamental solution $f(x,y,z)$ as follows:
\BEA\label{FTparaxSol}
\widetilde{f}(k_x,k_y,z)=\widetilde{f}(\bK,z) &= \frac{1}{2 \pi} \int\!\!\!\!\int f(x,y,z) \exp\left[- i  (x k_x + y k_y)\right] \rmd x \rmd y \nonumber \\
&=  i  \sqrt{\frac{L}{\pi k}} \exp\left[ -\frac{ i }{2 k }\left( k_x^2 + k_y^2\right) \left( z -  i  L\right) \right],
\EEA
which is normalized as well.
%
%

\section{Paraxial wave equation: quantum mechanics notation}
It is well known that paraxial classical optics is formally equivalent to two-dimensional quantum mechanics \cite{stoler,meron,kogelnik, 
nienhuis} and thus, we shall rewrite all aforementioned results in quantum-mechanical notation. However, 
although we are about to use the mathematical formalism of quantum mechanics to describe the propagation of a paraxial beam, it is worth stressing once more that the physics underneath is purely classical, and no quantum character of the electromagnetic field is taken into account at this level of description.

Given a generic function $\psi(\bor)=\psi(\bR,z)$ and its two-dimensional Fourier transform $\widetilde{\psi}(\bK,z)$ we define the position and momentum eigenkets $\ket{\bR}$ and $\ket{\bK}$, respectively, via the relations $\psi (\bR,z) = \brak{\bR}{ \psi(z)}$ and $\wpsi (\bK,z) = \brak{\bK}{ \psi(z)}$.
%
%
We assume that both $\ket{\bR}$ and $\ket{\bK}$ form a complete and orthonormal basis. Explicitly, for $\ket{\bR}$ this means:
%
%
\BEA
\int \ket{\bR}\!\bra{\bR} \rmd^2 R &= \bhone \qquad \mathrm{and} \qquad \brak{\bR}{\bR'} &= \delta\left(\bR - \bR' \right),\label{completenessA} 
\EEA
%
%
where with $\bhone$ we have denoted the identity operator. The two-dimensional Fourier transform fixes the value of $\brak{\bR}{\bK}$:
\BEA\label{PlaneWave1}
\wpsi( \bK,z )=\brak{\bK}{\psi(z)} &= \int \brak{\bK}{\bR} \brak{\bR}{\psi(z)} \rmd^2R \nonumber\\
&= \frac{1}{2 \pi} \int  e ^{- i  \bK \cdot \bR} \,\psi(\bR,z) \rmd x \rmd y,
\EEA
which implies $\brak{\bK}{\bR} =  e ^{- i  \bK \cdot \bR}/(2 \pi)$.
%
%
%
%
%

The position operator $\hbR = \{\hx, \hy \} = \{\hx_1, \hx_2 \}$ and momentum operator $\hbK= \{\hk_x, \hk_y \}= \{\hk_1, \hk_2 \}$ are defined via the eigenvalue equations $\hbR \ket{\bR'} = \bR' \ket{\bR'}$ and $\hbK \ket{\bK'} = \bK' \ket{\bK'}$.
%
%
%
%
They
fulfill the canonical commutation relations, i. e. $\bigl[ \hx_\alpha, \hk_\beta \bigr] =  i  \delta_{\alpha \beta}$ and $\bigl[ \hx_\alpha, \hx_\beta \bigr] = \bigl[ \hk_\alpha, \hk_\beta \bigr] =0$ for $\alpha, \beta \in \{1,2 \}$.
%
%
%
%

Finally, the position operator in the momentum basis and the momentum operator in the position basis are represented by
%
%
\numparts
\begin{eqnarray}\label{momOp}
\bra{\bK''} \hbR \ket{\bK'} = & \; i  \frac{\partial}{\partial \bK''} \delta( \bK' {-} \bK'') , \\
\bra{\bR''} \hbK \ket{\bR'} = & \; \frac{1}{ i}  \frac{\partial}{\partial \bR''} \delta( \bR' {-} \bR'') ,
\end{eqnarray}
\endnumparts
%
%
respectively, where $\frac{\partial}{\partial \bV}$ is a shorthand notation for the vector $\{ \frac{\partial}{\partial v_1},  \frac{\partial}{\partial v_2} \}$.

In this quantum formalism, the paraxial wave equation \eref{paraxEq1} can be rewritten as:
\BEA\label{paraxEq2}
0&=\left[ \frac{1}{2 k}\!\left(\frac{\partial^2}{\partial x^2} + \frac{\partial^2}{\partial y^2} \right)\!+   i  \frac{\partial}{\partial z} \right]f(x,y,z)
= \bra{\bR}\! \left( i \frac{\partial}{\partial z}  -\frac{1}{2k} \hbK^2  \right)\! \ket {f(z)},
\EEA
which is equivalent to the Schr\"{o}dinger equation for the free propagation of a quantum particle of ``mass'' $k$:
\BEA\label{paraxEq3}
 i \frac{\partial}{\partial z}\ket {f(z)} =   \frac{1}{2k} \hbK^2   \ket {f(z)}.
\EEA
The formal solution of this equation can thus be written in an operator form as:
\BEA\label{paraxProp}
\ket {f(z)} = \exp \left[ -\frac{ i }{2 k}  \hbK^2 \,(z - z_0) \right] \ket {f(z_0)},
\EEA
where $z_0$ is an arbitrary real constant \cite{stoler}.
%

\section{Beam shifts: from classical to quantum formalism}
%
%

\subsection{Quantum-mechanical representation of a reflection process}

Owing to the one-to-one correspondence between the paraxial wave equation and the Schr\"odinger equation, the electric field $\bE(\bor,t)$ of a paraxial beam can be described in a way formally equivalent to the wave function of a nonrelativistic quantum-mechanical particle with spin $1/2$. Let $\be_1 = \{1,0 \}$ and $\be_2 = \{0,1\}$ be two unit vectors that span the transverse plane perpendicular to the beam propagation axis $z$.
Then, we can write${}^2$\footnote[0]{$\!\!\!\!\!\!\!{}^2$\;\;Please note that henceforth we are working in ``natural'' units, where $k = 2 \pi/\lambda=1$ and $\mathrm{[length] = [wavenumber]} = 1$.} 
\BEA\label{Efieldparax0}
\bE(\bor,t) \propto   e ^{ i  (z - c t)} \sum_{\alpha =1}^2 \be_\alpha  \bra{\alpha}\brak{\bR}{\Psi(z)} ,
\EEA
where $\ket{\Psi(z)} = \ket{\psi(z)} \ket{A}$, with $\ket{A}$ being a two-component spinor representing the polarization of the beam, i.e. $\ket{A} \equiv a_1 \be_1+a_2\be_2=  \{	a_1, a_2 \}$ and $\abs{a_1}^2 + \abs{a_2}^2 =1$. Therefore, 
%
\BEA\label{Efieldparax1}
\bra{\alpha}\brak{\bR}{\Psi(z)} &= \brak{\bR}{\psi(z)} \brak{\alpha}{A} = \psi(\bR,z) a_\alpha,
\EEA
with $\psi(\bR,z)$ denoting a solution of the paraxial wave equation \eref{paraxEq1} and $\ket{\alpha=1} \equiv \be_1=\{1,0\}$ as well as $\ket{\alpha=2} \equiv \be_2=\{0,1\}$. 

The reflection process may be described by means of the scattering (entangling) operator
\BEA
\hat{S} = \sum_{\alpha=1}^2 \hat{M}_{(\alpha)} \otimes \hat{P}_{(\alpha)},
\EEA
where $\hat{P}_{(\alpha)}$ and $\hat{M}_{(\alpha)}$ are the polarization and the mode scattering
operators associated to the considered reflection process, respectively. These operators are defined as
\BEA
\hat{P}_{(\alpha)} = r_\alpha(\theta) |\alpha \rangle\langle \alpha|,
\EEA
with $r_1(\theta)$ and $r_2(\theta)$  being the Fresnel reflection coefficients evaluated at the incident angle $\theta$
(note the correspondence $1\, \Leftrightarrow \, p$-polarization and $2\, \Leftrightarrow \, s$-polarization) and
\BEA\label{scattering_mode}
\langle \bR |  \hat{M}_{(\alpha)}  | \psi(z) \rangle = \psi\left( -x + X_\alpha, y - Y_\alpha; z\right), 
\EEA
where the minus sign in the $x$-dependence of the shifted distribution in (\ref{scattering_mode}) is due to the parity inversion caused by reflection as seen from the reflected-beam reference frame.
We dedicate to this operation the ``bar'' symbol: if $\bV = \{v_x, v_y \}$ then $\bar{\bV} = \{-v_x, v_y \}$. The vector state representing the electric field after reflection can be thus written as
\BEA
 \hat{S} | \Psi(z) \rangle = \sum_{\alpha=1}^{2} \hat{M}_{(\alpha)}  | \psi(z) \rangle \hat{P}_{(\alpha)} |A \rangle =
\sum_{\alpha=1}^{2} a_\alpha r_\alpha(\theta)\hat{M}_{(\alpha)}  | \psi(z) \rangle  | \alpha \rangle  .
\EEA
The four dimensionless quantities $X_\alpha$ and $Y_\alpha$ with $\alpha\in\{1,2\}$ that appear in \eref{scattering_mode} are the \emph{complex} shifts, whose explicit forms are given by \cite{njp_gh_if}:
\numparts\label{ComplexShifts}
\BEA
X_{\, {1}} &= - i  \frac{\partial \ln r_{{1}}}{\partial \theta},   \qquad\qquad
Y_{\, {1}} &=   i  \frac{a_{ {2}}}{a_{\, {1}}}\left( 1 + \frac{r_{ {2}}}{r_{{1}}}\right) \cot \theta,  \label{Yp} \\
X_{ {2}} &=  - i  \frac{\partial \ln r_{ {2}}}{\partial \theta},  \qquad\qquad
Y_{ {2}} &=  - i  \frac{a_{\, {1}}}{a_{ {2}}}\left( 1 + \frac{r_{ {1}}}{r_{ {2}}}\right) \cot \theta. \label{Yo}
\EEA
\endnumparts
Finally, the vector wave function $\boldsymbol{\Psi}(\bR,z)=\sum_{\alpha =1}^2 \be_\alpha  \bra{\alpha}\bra{\bR}\hat{S}\ket{\Psi(z)}$ of a beam reflected by a plane surface may be written, with respect to a Cartesian reference frame attached to the reflected beam itself, as:
\BEA
\label{reflected_wave_func}
\boldsymbol{\Psi}(\bR,z) = \sum_{\alpha =1}^2 \be_\alpha  a_{\, \alpha} r_{\alpha}(\theta)\, \psi(\obR  - \obR_\alpha,z),
 \label{AnaSigRef2}
\EEA
where $\bR_\alpha = \{X_\alpha, Y_\alpha \}$. This result coincides with the according expressions obtained in \cite{merano,njp_gh_if}.
The shifted function $\psi(\obR  - \obR_\alpha,z) $ can be expanded in a Taylor series as follows:
\BEA
\label{shifted_func_taylor}
\psi(\obR  - \obR_\alpha,z) 
&\cong 
\psi(\obR,z)  - \bR_\alpha \cdot \frac{\partial}{\partial \bR}\psi(\obR,z) + \dots
\EEA
Using this result in \eref{AnaSigRef2} yields
\BEA\label{AnaSigRef3}
\boldsymbol{\Psi}(\bR,z) &\cong \sum_{\alpha =1}^2 \be_\alpha  a_{\alpha} r_{\alpha}(\theta)\!\left[ 1 - X_\alpha \frac{\partial}{\partial x} - Y_\alpha \frac{\partial}{\partial y} \right]\psi(\obR ,z).
\EEA
Here, the first term  gives the geometric optics contribution to the reflected beam (it may be simply named the ``Fresnel term''), while the second and the third terms are responsible for the GH and the IF shifts, respectively.


To put \eref{AnaSigRef3} in a fully quantum-mechanical form, we need to introduce the three $2\times2$ matrices
\numparts
\BEA
\mathsf{F} \equiv  & \, \left[
\begin{array}{cc}
	r_1(\theta)  & 0 \\
	0 & 	r_2(\theta)
\end{array}
\right], \label{FresnelMat} \\
\mathsf{X} \equiv  & \, \left[
\begin{array}{cc}
	-  i  \frac{\partial \ln r_1}{\partial \theta}  & 0 \\
	0 & 	-  i  \frac{\partial \ln r_2}{\partial \theta}
\end{array}
\right], \label{GHMat} \\
 \mathsf{Y} \equiv  & \, \left[
\begin{array}{cc}
	0  &  i  \left( 1 + \frac{r_1}{r_2} \right) \cot \theta \\
	- i  \left( 1 + \frac{r_2}{r_1} \right) \cot \theta & 	0
\end{array}
\right] , \label{IFMat}
\EEA
\endnumparts
that represent the Fresnel reflection ($\mathsf{F}$), the GH ($\mathsf{X}$) and the IF ($\mathsf{Y}$) shifts, respectively. Then it is not difficult to see by means of a straightforward calculation that 
\BEA\label{AnaSigRef4}
\boldsymbol{\Psi}(\bR,z) &\cong \left[ \mathsf{I} - \frac{\partial}{\partial x} \,\mathsf{X}  - \frac{\partial}{\partial y} \,\mathsf{Y} \right]\cdot\mathsf{F} \cdot \left[
\begin{array}{c}
	a_1 \\
	a_2
\end{array}
\right]\psi(\obR ,z) \nonumber \\
&\cong \exp \left(- \hsfA  \cdot \frac{\partial}{\partial \bR} \right) \psi(\obR,z)\cdot \left[
\begin{array}{c}
	r_1 a_1 \\
r_2 a_2
\end{array}
\right],
\EEA
where $\mathsf{I}$ denotes the $2\times 2$ identity matrix and we have defined the matrix-valued ``spin operator'' vector $\hsfA = \{ \mathsf{X}, \mathsf{Y}\} \equiv \{ \mathsf{A}_1, \mathsf{A}_2\}$. Let $\hat{M}$ be the Hermitian and symmetric operator representing the mirror-symmetry reflection with respect to the  $x$ axis. By definition, it acts upon the position eigenket $\ket{\bR}=\ket{x,y}$ as follows: $\hat{M}\ket{x,y} = \ket{{-}x,y} \equiv \ket{\obR}$ and $\hat{M}^2\ket{x,y} = \hat{M}\ket{{-}x,y}= \ket{x,y}$, namely $\hat{M}^2 = \bhone$. Then we can write
\BEA\label{MirrSym}
\psi(\obR,z) &=\brak{\obR}{\psi(z)} = \bra{\bR}\hat{M}\ket{\psi(z)} \equiv \brak{\bR}{\opsi(z)}.
\EEA
With the help of \eref{momOp} and \eref{Efieldparax1} it is straightforward to write \eref{AnaSigRef4} as
\BEA\label{AnaSigRef5}
\boldsymbol{\Psi}(\bR,z) = \brak{\bR}{\Psi(z)} =  \bra{\bR}\exp \bigl(- i  \hsfA  \cdot \hbK \bigr) \ket{\opsi(z)}\ket{ A_\mathsf{F}}
,
\EEA
where $\ket{ A_\mathsf{F}} \equiv \mathsf{F}\ket{ A}=\{ r_1 a_1, r_2 a_2 \}$. The operator $\hat{H}_I = \hsfA  \cdot \hbK$ is exactly what Hosten and Kwiat \cite{hosten} call the ``interaction Hamiltonian'' that couples the momentum of the
meter to the ``spin observable'' $ \hsfA$. However, it should be noticed that this ``Hamiltonian'' is not Hermitian because $ \hsfA \neq \hsfA^\dagger$. Therefore, the operator $ \hsfA$ does not really correspond to an observable. As a consequence of this, the operator $\exp (-  i  \hsfA  \cdot \hbK )$ is not unitary.
 We shall find later that only the Hermitian combinations $  \hsfA +   \hsfA^\dagger$ and $ - i ( \hsfA -  \hsfA^\dagger)$ are indeed observables and yield to the measurable spatial and angular shifts, respectively.
 
Finally, by using  \eref{paraxProp}, \eref{AnaSigRef5} can be recast in a fully quantum-mechanical form as:
\BEA\label{AnaSigRef6}
\ket{\Psi(z)} &= \exp \bigl(-  i  \hsfA  \cdot \hbK \bigr) \ket{\opsi(z)}\ket{ A_\mathsf{F}} \nonumber \\
&= \exp \bigl(-  i  \hsfA  \cdot \hbK \bigr) \exp \left( -\frac{ i }{2 }  \hbK^2 \, z \right) \ket{\opsi(0)}\ket{ A_\mathsf{F}}.
\EEA
It should be noticed that the 
free propagator $\exp ( - i   \hbK^2 \, z/{2 } )$ and the interaction operator $\exp(-  i  \hsfA  \cdot \hbK )$ do commute.
%
%
%

\subsection{Ordinary (not weak)  beam shifts}
%

Beam shifts are quantified by the displacement of the centroid of the beam distribution after reflection with respect to the centroid of the reflected beam according to geometric optics. Hence, we calculate the expectation value of the position operator $\hbR$ 
in the reference frame attached to the reflected beam, namely
\BEA\label{Shift1}
\langle \bR \rangle(z) = \frac{\bra{\Psi(z)} \hbR \ket{\Psi(z)}}{\brak{\Psi(z)}{\Psi(z)}},
\EEA
where $\ket{\Psi(z)}$ has been defined for the reflected beam in \eref{AnaSigRef6}. Using \eref{AnaSigRef6}, the denominator of the expression above gives 
\BEA\label{Denom}
\brak{\Psi(z)}{\Psi(z)} &= \bra{ A_\mathsf{F}}\bra{\opsi(0)}   e ^{ \frac{ i }{2 }  \hbK^2 \, z } e ^{ i  \hsfA^\dagger  \cdot \hbK }   e ^{- i  \hsfA \cdot \hbK }  e ^{- \frac{ i }{2 }  \hbK^2 \, z }\ket{\opsi(0)}\ket{ A_\mathsf{F}} \nonumber \\
%
&\cong 
%
\bra{ A_\mathsf{F}}\bra{\opsi(0)} \left[ 1 -  i  \left( \hsfA  - \hsfA^\dagger \right) \cdot \hbK + \ldots \right] \ket{\opsi(0)}\ket{ A_\mathsf{F}} \nonumber \\
%
%
&= \brak{ A_\mathsf{F}}{ A_\mathsf{F}}  \brak{\opsi(0)}{\opsi(0)}  + \mathrm{quadratic~terms~in~}\hsfA,
\EEA
since $\bra{\opsi(0)} \hbK \ket{\opsi(0)} =0$ because it amounts to the angular shift of the input beam which is, by definition, zero. If we assume that the input wave function is normalized, then $ \brak{\opsi(0)}{\opsi(0)} =\bra{\psi(0)}\hat{M}^2\ket{\psi(0)}=\brak{\psi(0)}{\psi(0)}=1$ and we can rewrite
\BEA\label{Shift2}
\brak{\Psi(z)}{\Psi(z)} &\cong  \brak{ A_\mathsf{F}}{ A_\mathsf{F}} = \abs{r_1 a_1 }^2 + \abs{r_2 a_2 }^2.
\EEA
Equation \eref{Denom} clearly illustrate a result that we already know from conventional calculations, namely that the perturbative corrections to the denominator start at second order \cite{REF_3}.

The numerator of  \eref{Shift1} reads as:
\BEA\label{Num}
\bra{\Psi(z)}\hbR \ket{\Psi(z)} = \bra{ A_\mathsf{F}}\bra{\opsi(z)}  e ^{ i  \hsfA^\dagger  \cdot \hbK }
\hbR
 e ^{- i  \hsfA \cdot \hbK }\ket{\opsi(z)}\ket{ A_\mathsf{F}},
\EEA
and we can evaluate it first by noticing that
\numparts
\BEA
\label{Num1a}
 e ^{ i  \hsfA^\dagger  \cdot \hbK }  \hbR  e ^{- i  \hsfA \cdot \hbK }  &=  e ^{ i  \hsfA^\dagger  \cdot \hbK }  e ^{- i  \hsfA \cdot \hbK } \left(  e ^{ i  \hsfA \cdot \hbK }\hbR  e ^{- i  \hsfA \cdot \hbK } \right)\\
\label{Num1b}
&= \left(  e ^{ i  \hsfA^\dagger \cdot \hbK }\hbR  e ^{- i  \hsfA^\dagger \cdot \hbK } \right)  e ^{ i  \hsfA^\dagger  \cdot \hbK }  e ^{- i  \hsfA \cdot \hbK }.
\EEA
\endnumparts
By using the Baker-Campbell-Hausdorff lemma \cite{merzbacher} and the canonical commutation relation of $\hbR$ and $\hbK$ it is straightforward to calculate
\BEA\label{Num2}
 e ^{ i  \hsfA \cdot \hbK }\hbR  e ^{- i  \hsfA \cdot \hbK } = \hbR + \hsfA\qquad\mathrm{and}\qquad e ^{ i  \hsfA^\dagger \cdot \hbK }\hbR  e ^{- i  \hsfA^\dagger \cdot \hbK } = \hbR + \hsfA^\dagger.
\EEA
Moreover, from the calculation of the denominator we already know that
\BEA\label{Num3}
  e ^{ i  \hsfA^\dagger  \cdot \hbK }  e ^{- i  \hsfA \cdot \hbK } \cong  1 -  i  \left( \hsfA  - \hsfA^\dagger \right) \cdot \hbK + \ldots \; ,
\EEA
Inserting \eref{Num2} and \eref{Num3} into \eref{Num1a} we obtain
\numparts
\BEA\label{Num4a}
 e ^{ i  \hsfA^\dagger  \cdot \hbK }  \hbR \,  e ^{- i  \hsfA \cdot \hbK } &\cong \left[ 1 -  i  \left( \hsfA  - \hsfA^\dagger \right) \cdot \hbK + \ldots \right] \left( \hbR + \hsfA \right) \nonumber \\
&\cong \hbR + \hsfA -  i \left[ \left( \hsfA  - \hsfA^\dagger \right) \cdot \hbK \right] \hbR + \ldots.
\EEA
and, similarly, from \eref{Num1b} we attain
\BEA\label{Num4b}
 e ^{ i  \hsfA^\dagger  \cdot \hbK }  \hbR \,  e ^{- i  \hsfA \cdot \hbK } &\cong  \left( \hbR + \hsfA^\dagger \right)\left[ 1 -  i  \left( \hsfA  - \hsfA^\dagger \right) \cdot \hbK + \ldots \right] \nonumber \\
&\cong \hbR + \hsfA^\dagger -  i  \hbR\left[ \left( \hsfA  - \hsfA^\dagger \right) \cdot \hbK \right] + \ldots ,
\EEA
\endnumparts
where quadratic and higher order terms in $\hsfA$ and $\hsfA^\dagger$ have been discarded in the last lines since the shifts $X_\alpha$ and $Y_\alpha$ with $\alpha\in\{1,2\}$ are supposed to be small. The approximations \eref{Num4a} and \eref{Num4b} are not Hermitian, although the left sides of these equations are. 
To obtain a Hermitian quantity for the approximation of the left hand side of \eref{Num1a}
, we take a symmetric combination of \eref{Num4a} and \eref{Num4b}
. By substituting this into the numerator of \eref{Num} we find:
%
%
\BEA\label{Num5}
\bra{\Psi(z)}\hbR \ket{\Psi(z)} 
%
&\cong \bra{\opsi(z)}\hbR  \ket{\opsi(z)}\brak{ A_\mathsf{F}}{ A_\mathsf{F}}\nonumber\\
&\quad + \frac{1}{2}\brak{\opsi(z)}{\opsi(z)}\bra{ A_\mathsf{F}}\!\left(\hsfA+\hsfA^\dagger\right)\!\ket{ A_\mathsf{F}} \nonumber \\
&\quad- \frac{ i }{2} \bra{\opsi(z)} \!\left(\hbK\hbR + \hbR\hbK\right)\! \ket{\opsi(z)}  {\cdot}  \bra{A_\mathsf{F}} \!\left( \hsfA  - \hsfA^\dagger \right)\! \ket{ A_\mathsf{F}}\nonumber \\
%
&=  \mathrm{Re}\bra{ A_\mathsf{F}}\hsfA\ket{ A_\mathsf{F}}\nonumber \\
&\quad + \bra{\opsi(z)}\!\left(\hbK\hbR + \hbR\hbK\right)\!\ket{\opsi(z)} {\cdot} \mathrm{Im}\bra{A_\mathsf{F}}\hsfA\ket{ A_\mathsf{F}}  
%
%
\EEA
where $\brak{\opsi(z)}{\opsi(z)}=\brak{\psi(z)}{\psi(z)}=1$ and $\bra{\opsi(z)}\hbR  \ket{\opsi(z)}=\bra{\opsi(0)}\hbR+z\hbK \ket{\opsi(0)} =0$ has been used. The latter equality is due to the result 
\BEA
e^{\frac{i}{2}\hbK^2 z}\hbR e^{-\frac{i}{2}\hbK^2 z}=\hbR+z\hbK,
\EEA
and the fact that spatial and angular shift of the input beam is zero by definition. Note that in \eref{Num5}, the dyadic operator $\hbK\hbR + \hbR\hbK$ can be represented by a $2 \times 2$ matrix by recalling that $[\hbK \hbR + \hbR \hbK]_{\alpha \beta} = \hk_\alpha \hx_\beta+\hx_\alpha \hk_\beta$:
%
\BEA\label{dyadic}
\hbK  \hbR + \hbR  \hbK = \left[
\begin{array}{cc}
	\hk_x \hx + \hx \hk_x \; & \; \hx \hk_y + \hy \hk_x \\
  \hx \hk_y + \hy \hk_x \; & \; \hk_y \hy + \hy \hk_y
\end{array} \right].
\EEA
This matrix formulation is equivalent to equations (\ref{reflected_wave_func}) and (\ref{shifted_func_taylor}) in \cite{njp_gh_if} when calculating the shift of an OAM beam. This can be shown explicitly by calculating the expectation value
\BEA
\bra{\opsi(z)}\hbR\hbK\ket{\opsi(z)}&=\int\rmd^2 R\int\rmd^2 R'\bra{\opsi(z)}\hbR\ket{\bR'}\!\bra{\bR'}\hbK\ket{\bR}\!\brak{\bR}{\opsi(z)}\nonumber\\
&=- i \int\rmd^2 R\,\psi^*(\obR,z)\bR\frac{\partial}{\partial \bR}\psi(\obR,z)
\EEA
and therefore
\BEA
\label{comparison}
\bra{\opsi(z)}\!\left(\hbK\hbR + \hbR\hbK\right)\!\ket{\opsi(z)}&=2\mathrm{Re}\bra{\opsi(z)}\hbR\hbK\ket{\opsi(z)}\nonumber\\
&=2\mathrm{Im}\int\rmd^2 R\,\psi^*(\obR,z)\bR\frac{\partial}{\partial \bR}\psi(\obR,z).
\EEA
Thus, it is not by chance that in \cite{njp_gh_if} the off diagonal matrix entries calculated for an OAM beam are proportional to the angular momentum carried by the beam itself. It is clear from \eref{comparison} that the off-diagonal elements of $\hbK  \hbR + \hbR \hbK$, namely $\hx \hk_y + \hy \hk_x$, are proportional to the $z$ component of the angular momentum operator.


The $z$-dependence in the term $\bra{\opsi(z)} \hbK  \hbR +  \hbR  \hbK \ket{\opsi(z)}$ may be explicitly calculated:
\BEA\label{Num6}
\bra{\opsi(z)} \hk_\alpha \hx_\beta \ket{\opsi(z)} &= \bra{\opsi(0)} e ^{ \frac{ i }{2 }  \hbK^2 \, z }  \hk_\alpha \hx_\beta  e ^{ -\frac{ i }{2 }  \hbK^2 \, z } \ket{\opsi(0)} \nonumber \\
%
%
&= \bra{\opsi(0)}  \hk_\alpha   \bigl( \hx_\beta + z \, \hk_\beta \bigr)\ket{\opsi(0)} \nonumber \\
&= \bra{\opsi(0)}  \hk_\alpha  \hx_\beta \ket{\opsi(0)} + z \, \bra{\opsi(0)} \hk_\alpha  \hk_\beta \ket{\opsi(0)}.
\EEA
Similarly:
\BEA
\bra{\opsi(z)} \hx_\alpha \hk_\beta \ket{\opsi(z)} &= \bra{\opsi(0)}  \hx_\alpha  \hk_\beta \ket{\opsi(0)} + z \, \bra{\opsi(0)} \hk_\alpha  \hk_\beta \ket{\opsi(0)},
\EEA
where the Baker-Campbell-Hausdorff lemma has been used once again. Thus,  \eref{Num5} can be rewritten as
\BEA\label{Num7}
\bra{\Psi(z)}\hbR \ket{\Psi(z)} &\cong  \mathrm{Re}\bra{ A_\mathsf{F}}\hsfA\ket{ A_\mathsf{F}} +  \bra{\opsi(0)} \hbK  \hbR {+} \hbR  \hbK \ket{\opsi(0)} \cdot \mathrm{Im}\bra{A_\mathsf{F}} \hsfA \ket{ A_\mathsf{F}}\nonumber \\
& \quad + z \, \bra{\opsi(0)} 2\hbK  \hbK \ket{\opsi(0)}  \cdot  \mathrm{Im}\bra{A_\mathsf{F}} \hsfA \ket{ A_\mathsf{F}}.
\EEA
The angular shift $\Theta$ contribution comes from the term proportional to $z$, whereas the spatial shift $\Delta$ is independent of $z$ (see figure \ref{fig.shifts}). Therefore, the first line of \eref{Num7} gives the spatial shift. However, there are two contributions to the spatial shift. The first term of the first line of \eref{Num7}, proportional to the real part of $\hsfA$, gives a spatial shift that only depends on the polarization properties of the beam and (throught the matrix $\mathsf{F}$) on the properties of the surface. Conversely, the second term on the same line, proportional to the imaginary part of $\hsfA$, depends also on the spatial properties of the beam and yields the spatial-vs-angular shift mixing occurring, for example, for OAM beams. Such OAM-induced beam shifts were predicted theoretically by Fedoseyev \cite{Fedoseyev2001,Fedoseyev2008} and by Bliokh and coworkers \cite{BSK} and in the case of the IF shift observed experimentally by Dasgupta and coworkers \cite{Dasgupta}.
Finally, the second line of the equation above gives the angular shift ($z$-dependent part of the total shift). We find that it is always proportional to the angular spread of the beam because it amounts to the momentum self-correlation matrix $\hbK  \hbK$ whose diagonal elements give indeed the angular spread of the incident beam. Thus, the beams angular aperture is proportional to the variance of the transverse component of the $k$-vector.

To illustrate in greater detail \eref{Num7} let us calculate it for the specific case of an input fundamental Gaussian beam of the form \eref{paraxSolFund}, namely for:
\BEA\label{Num8}
\brak{\bR}{\opsi(0)} = \frac{ i }{\sqrt{\pi L }}\exp\left(-\frac{x^2 + y^2}{2L}\right),
\EEA
and
\BEA\label{Num9}
\brak{\bK}{\opsi(0)} =  i  \sqrt{\frac{L}{\pi}} \exp\left(-\frac{k_x^2 + k_y^2}{2/L}\right).
\EEA
A straightforward calculation furnishes:
\numparts
\BEA\label{Num10}
\bra{\opsi(0)} \hbK  \hbR + \hbR \hbK \ket{\opsi(0)} = 0\qquad \mathrm{and} \qquad\bra{\opsi(0)} \hbK  \hbK \ket{\opsi(0)} = \frac{1}{2 L} \bone.
\EEA
\endnumparts
Finally, by using these results \eref{Num7} reduces to
\BEA\label{Num11}
\bra{\Psi(z)}\hbR \ket{\Psi(z)} &= \mathrm{Re} \bra{ A_\mathsf{F}}\hsfA\ket{ A_\mathsf{F}} + \frac{z}{L} \mathrm{Im}\bra{A_\mathsf{F}}\hsfA\ket{ A_\mathsf{F}}.
\EEA
This clear result beautifully illustrates how the real and the imaginary part of the interaction operator $\hsfA $ yield to the spatial and the angular shifts, respectively. At the end of the day, gathering all the result we can write
\BEA\label{Num12}
\frac{\bra{\Psi(z)}\hbR \ket{\Psi(z)}}{\brak{\Psi(z)}{\Psi(z)}}
=   \mathrm{Re}  \frac{\bra{ A_\mathsf{F}} \hsfA \ket{ A_\mathsf{F}} }{\brak{ A_\mathsf{F}}{ A_\mathsf{F}}}  + \frac{z}{L} \, \mathrm{Im}  \frac{\bra{ A_\mathsf{F}} \hsfA \ket{ A_\mathsf{F}} }{\brak{ A_\mathsf{F}}{ A_\mathsf{F}}},
\EEA
which, in terms of the matrices $\mathsf{X}$ and $\mathsf{Y}$ given by equations \eref{GHMat} and \eref{IFMat}, respectively, may be rewritten as:
\numparts\label{Num13}
\BEA
\Delta_\mathrm{GH}
&=  \mathrm{Re}  \frac{\bra{ A_\mathsf{F}} \mathsf{X} \ket{ A_\mathsf{F}} }{\brak{ A_\mathsf{F}}{ A_\mathsf{F}}},   \qquad\qquad \Theta_\mathrm{GH}
&=\frac{1}{L} \, \mathrm{Im}  \frac{\bra{ A_\mathsf{F}}  \mathsf{X} \ket{ A_\mathsf{F}} }{\brak{ A_\mathsf{F}}{ A_\mathsf{F}}}, \\
\Delta_\mathrm{IF} &= \mathrm{Re}  \frac{\bra{ A_\mathsf{F}} \mathsf{Y} \ket{ A_\mathsf{F}} }{\brak{ A_\mathsf{F}}{ A_\mathsf{F}}},   \qquad\qquad \Theta_\mathrm{IF} &= \frac{1}{L} \, \mathrm{Im}  \frac{\bra{ A_\mathsf{F}}  \mathsf{Y} \ket{ A_\mathsf{F}} }{\brak{ A_\mathsf{F}}{ A_\mathsf{F}}}.
\EEA
\endnumparts
These expressions are fully coincident with the ones obtained by means of ordinary (classical) calculations for an input fundamental Gaussian beam \cite{shift_overview}.
%
%

\subsection{Weak measurements}
Theory \cite{aiello2008,gotte1,gotte2,jozsa} points out a close connection between beam shifts and weak measurements \cite{aharonov,ritchie}. In experiments \cite{hosten, qin} weak measurements are frequently applied to enhance and thus observe beam shift effects. For this reason, we will extend our formalism now to weak measurements. To begin with, let us rewrite \eref{AnaSigRef6} as:
\BEA\label{Weak1}
\ket{\Psi(z)} &= \exp \bigl(-  i  \hsfA  \cdot \hbK \bigr) \ket{\opsi(z)}\ket{ A_\mathsf{F}} ,
\EEA
which describes the beam up to the detector surface. Now, imagine to put in front of the detector a polarizer oriented along the direction $\ket{B} = b_1 \be_1 + b_2 \be_2$, with $\abs{b_1}^2 + \abs{b_2}^2 =1$. As consequence, the polarization of the beam will be projected along this direction and the resulting state will be:
\BEA\label{Weak2}
\ket{\Psi(z)} \rightarrow   \ket{ B }\brak{ B }{\Psi(z)} \equiv \ket{\psi_B(z)}\ket{ B } ,
\EEA
where $\ket{ B } \equiv \{b_1, b_2 \}$ and
\BEA\label{Weak3}
\ket{\psi_{B} (z)} =  \bra{ B }\exp \bigl(-  i  \hsfA  \cdot \hbK \bigr) \ket{\opsi(z)}\ket{ A_\mathsf{F}} .
\EEA
As usual, in the hypothesis of weak perturbation we can expand the exponential to obtain
\BEA\label{Weak4}
\ket{\psi_{B} (z)} &\cong  \bra{ B } \bigl(1-  i  \hsfA  \cdot \hbK  + \dots \bigr) \ket{\opsi(z)}\ket{ A_\mathsf{F}} \nonumber \\
%
%
&= \brak{ B }{ A_\mathsf{F}}\left[ \ket{\opsi(z)} -  i  \frac{ \bra{ B } \hsfA \ket{ A_\mathsf{F}}}{\brak{ B }{ A_\mathsf{F}}} \cdot \hbK \ket{\opsi(z)}  + \dots  \right] \nonumber \\
&\cong \brak{ B }{ A_\mathsf{F}}\exp \bigl( -  i  \sfA_w \cdot \hbK \bigr) \ket{\opsi(z)},
\EEA
where we have defined the vector-valued weak value
\BEA\label{Weak5}
\sfA_w =  \frac{ \bra{ B } \hsfA \ket{ A_\mathsf{F}}}{\brak{ B }{ A_\mathsf{F}}},
\EEA
with a post-selection probability $P_\mathrm{ps} = \abs{\brak{ B }{ A_\mathsf{F}}}^2/\brak{ A_\mathsf{F} }{ A_\mathsf{F}}$. Since $\sfA_w  \in \mathbb{C}$ this quantity is not directly observable. However, it is simply related to the spatial and angular beam shifts as it can be seen by calculating explicitly the post-selected wave function in the position representation:
\BEA\label{Weak6}
\brak{\bR}{\psi_{B}(z)} &= \brak{ B }{ A_\mathsf{F}}\bra{\bR}\exp \bigl( -  i  \sfA_w \cdot \hbK \bigr) \ket{\opsi(z)}\nonumber \\
&= \brak{ B }{ A_\mathsf{F}}\int \brak{\bR}{\bK}\bra{\bK}\exp \bigl( -  i  \sfA_w \cdot \hbK \bigr) \ket{\opsi(z)} \rmd^2 K \nonumber \\
&= \frac{\brak{ B }{ A_\mathsf{F}}}{2 \pi} \int \exp \bigl(  i \bR \cdot \bK \bigr) \exp \bigl( -  i  \sfA_w \cdot \bK \bigr)\brak{\bK}{\opsi(z)} \rmd^2 K \nonumber \\
&= \frac{\brak{ B }{ A_\mathsf{F}}}{2 \pi} \int \exp \bigl[i \bK \cdot \bigl(  \bR - \sfA_w  \bigr) \bigr] \wpsi(\bar{\bK},z)
\rmd^2 K \nonumber \\
&= \brak{ B }{ A_\mathsf{F}} \, \psi(\obR - \bsfA_w,z).
\EEA
At this point one may proceed in the usual manner calculating the centroid of the shifted distribution $ |\psi(\obR - \bsfA_w, z)|^2$.
Alternatively, we can start from \eref{Weak4} and calculate directly $\langle \bR \rangle(z) $ from \eref{Shift1}. By proceeding in exactly the same manner as before, first we find $\brak{\psi_{B}(z)}{\psi_{B}(z)} = \abs{\brak{B}{A_\mathsf{F}}}^2 = \abs{\bra{B}\mathsf{F}\ket{A}}^2$. Then, we calculate the corresponding of \eref{Num7} which yield to our final result:
\BEA\label{Weak7}
\frac{\bra{\psi_{B}(z)}\hbR \ket{\psi_{B}(z)}}{\brak{\psi_{B}(z)}{\psi_{B}(z)}} &= \frac{\sfA_w  + \sfA_w ^\dagger}{2}+  \bra{\opsi(0)} \hbK  \hbR +  \hbR  \hbK  \ket{\opsi(0)}  \cdot  \frac{\sfA_w  - \sfA_w ^\dagger}{2 i}\nonumber \\
&\quad + z \, \bra{\opsi(0)}2 \hbK  \hbK \ket{\opsi(0)}  \cdot  \frac{\sfA_w  - \sfA_w ^\dagger}{2 i}.
\EEA
Explicitly, for an input fundamental Gaussian beam, the post-selected GH and IF shifts are thus:
\numparts\label{Weak8}
\BEA
\Delta_\mathrm{GH}^\mathrm{ps}&= \mathrm{Re}  \frac{\bra{ B } \mathsf{X} \ket{ A_\mathsf{F}} }{\brak{ B }{ A_\mathsf{F}}}  \qquad\qquad \Theta_\mathrm{GH}^\mathrm{ps}&= \frac{1}{L} \, \mathrm{Im}  \frac{\bra{ B }  \mathsf{X}\ket{ A_\mathsf{F}} }{\brak{ B }{ A_\mathsf{F}}}, \\
\Delta_\mathrm{IF}^\mathrm{ps}&= \mathrm{Re}  \frac{\bra{ B } \mathsf{Y} \ket{ A_\mathsf{F}} }{\brak{ B }{ A_\mathsf{F}}}   \qquad\qquad \Theta_\mathrm{IF}^\mathrm{ps}&= \frac{1}{L} \, \mathrm{Im}  \frac{\bra{ B }  \mathsf{Y} \ket{ A_\mathsf{F}} }{\brak{ B }{ A_\mathsf{F}}}.
\EEA
\endnumparts
%
%

\section{Discussions}
In the previous section we have derived the following expressions for the (complex-valued) post-selection enhanced GH and IF shifts:
\numparts
\BEA
\label{enhanced_shifts}
X  \equiv \frac{\bra{ B} \mathsf{X} \ket{ A_\mathsf{F}} }{\brak{ B}{ A_\mathsf{F}}} &= - i \frac{b_1^* \frac{\partial \, r_1}{\partial \theta} a_1 + b_2^* \frac{\partial \, r_2}{\partial \theta} a_2}{ b_1^* r_1 a_1 + b_2^* r_2 a_2},  \label{Res61a}\\
Y \equiv  \frac{\bra{ B} \mathsf{Y} \ket{ A_\mathsf{F}} }{\brak{ B}{ A_\mathsf{F}}} &=  i (r_1 + r_2) \cot \theta \frac{b_1^* a_2 - b_2^* a_1 }{ b_1^* r_1 a_1 + b_2^* r_2 a_2}.\label{Res61b}
\EEA
\endnumparts
%
First of all, we notice that when either $a_1=0$ or $a_2=0$ one has:
\numparts
\BEA\label{Res63b}
X \big|_{a_1=0} &=  -\frac{ i  }{r_2}\frac{\partial \,  r_2}{\partial \theta} , \qquad \qquad \qquad 
X \big|_{a_2=0} &=   -\frac{ i }{r_1}\frac{\partial \,  r_1}{\partial \theta}, \label{GHsub} \\
Y \big|_{a_1=0} &=  i \frac{b_1^*}{b_2^*} \,\left(1 + \frac{r_1}{r_2} \right) \cot \theta, \qquad
Y \big|_{a_2=0} &=  - i \frac{b_2^*}{b_1^*} \,\left(1 + \frac{r_2}{r_1} \right) \cot \theta . \label{IFsub}
\EEA
\endnumparts
From \eref{GHsub} it follows that the ``ordinary'' GH shift, either spatial or angular, cannot be enhanced if one has either $s$ ($a_1=0$) or $p$ ($a_2=0$) input polarization. Vice versa, for the same kind of input states the IF shift can be enhanced. For example, consider the experiment by Hosten and Kwiat \cite{hosten}, 
where the input is polarized horizontally ($\ket{A}=\ket{H} \Leftrightarrow a_2=0$) and post-selected with linear polarization: $\ket{B} =-\sin \Delta \ket{H} + \cos \Delta \ket{V} = b_1 \ket{\alpha=1} + b_2 \ket{\alpha=2}$. In this case we obtain from the second equation of \eref{IFsub}:
\BEA\label{IFsub2}
\bigl.Y \bigr|_{a_2=0} =  i  \cot \Delta \,\left(1 + \frac{r_2}{r_1} \right) \cot \theta ,
\EEA
which grows indefinitely when $\Delta \rightarrow 0$. In the case of real $r_1$ and $r_2$, $Y$ becomes purely imaginary and the angular shift is therefore magnified, in agreement with Hosten and Kwiats experiment. Equations (\ref{Res61a}) and (\ref{Res61b}) are strictly valid until the denominators do not vanish. However, this is precisely what happens at Brewster's and null-reflection angles, as described with great detail in \cite{gotteOL}. In those conditions, (\ref{Res61a}) and (\ref{Res61b}) cease to be valid nearby the singularity and the higher terms disregarded in (\ref{Denom}), (\ref{Num4a}) and (\ref{Num4b}) contribute \cite{inpreparation}.


Next we discuss the question what is in general the best choice for the post selection state in these weak measurements. The expressions in \eref{Res61a} and \eref{Res61b} are clearly singular when $\brak{ B}{ A_\mathsf{F}}=0$ which occurs for
\BEA\label{Res64}
\ket{B}:&= \{-r_2^* a_2^* , r_1^* a_1^*\} \nonumber \\
&\equiv  \ket{A_\mathsf{F}^\perp} .
\EEA
So, in principle, one could choose $\ket{B} \simeq \ket{A_\mathsf{F}^\perp}$ in order to increase the magnitude of $X$, but at the same time keeping it finite. However, this is only part of the story because the choice of $\ket{B}$ also affects the value of the numerator in \eref{Res61a} and \eref{Res61b}. Thus, one could choose $\ket{B} = \mathsf{U} \ket{A_\mathsf{F}^\perp}$, where $\mathsf{U}$ is an arbitrary $2 \times 2 $ matrix (with $\mathsf{U} \neq \mathsf{I}$ in order to avoid the singularity) and use the 3 (three) real parameters from which $\mathsf{U}$ may depend, in order to maximize (numerically) $X$ and $Y$. This is far too complicated and so we use a single real parameter, say $\Delta \in [0, 2 \pi)$, and choose
\BEA\label{Res65}
\ket{B}= \cos \Delta \ket{A_\mathsf{F}^\perp}  - \sin \Delta \ket{A_\mathsf{F}},
\EEA
in order to have a huge enhancement for $\Delta \simeq 0$.
With this choice the post-selection probability
\BEA\label{Res66}
P_\mathrm{ps} = \frac{\;\abs{\brak{ B }{ A_\mathsf{F}}}^2}{\brak{ A_\mathsf{F} }{ A_\mathsf{F}}} =  \sin^2 \Delta,
\EEA
is independent from the angle of incidence of the beam and can be kept constant during the experiment. The intensity of the beam behind the post-selecting polarizer is simply equal to $I_\mathrm{ps} = P_\mathrm{ps}\brak{ A_\mathsf{F} }{ A_\mathsf{F}} = \sin^2 \Delta(|r_1 a_1 |^2 + |r_2 a_2 |^2)$.

From \eref{Res63b}, \eref{IFsub} and \eref{Res65} it follows that, in general, the enhanced shift can always be written as the sum of the ordinary shift plus an enhancement term:
\BEA\label{Res67}
S = \frac{\bra{ B} \mathsf{S} \ket{ A_\mathsf{F}} }{\brak{ B}{ A_\mathsf{F}}} &= \frac{\bra{ A_\mathsf{F}} \mathsf{S} \ket{ A_\mathsf{F}} }{\brak{ A_\mathsf{F}}{ A_\mathsf{F}}} + \frac{\; \; \brak{ B}{ A_\mathsf{F}^\perp} }{\brak{ B}{ A_\mathsf{F}}} \; \frac{\bra{ A_\mathsf{F}^\perp} \mathsf{S} \ket{ A_\mathsf{F}} }{\brak{ A_\mathsf{F}}{ A_\mathsf{F}}} \nonumber \\
&= \frac{\bra{ A_\mathsf{F}} \mathsf{S} \ket{ A_\mathsf{F}} }{\brak{ A_\mathsf{F}}{ A_\mathsf{F}}} - \cot \Delta \, \frac{\bra{ A_\mathsf{F}^\perp} \mathsf{S} \ket{ A_\mathsf{F}} }{\brak{ A_\mathsf{F}}{ A_\mathsf{F}}},
\EEA
where $S \in \{ X, Y\}$, $\mathsf{S} \in \{ \mathsf{X}, \mathsf{Y}\}$ and the completeness relation $\ket{A_\mathsf{F}}\!\bra{A_\mathsf{F}} + \ket{A_\mathsf{F}^\perp}\!\bra{A_\mathsf{F}^\perp} = \brak{A_\mathsf{F}}{A_\mathsf{F}}\bhone $ has been used. Note that since $\brak{ A_\mathsf{F}}{ A_\mathsf{F}}= \abs{r_1 a_1}^2 + \abs{r_2 a_2}^2=\brak{ A_\mathsf{F}^\perp}{ A_\mathsf{F}^\perp}$, the second term in (\ref{Res67}) can be written in the following more symmetric form:
\BEA\label{Res67B}
 \frac{\bra{ A_\mathsf{F}^\perp} \mathsf{S} \ket{ A_\mathsf{F}} }{\brak{ A_\mathsf{F}}{ A_\mathsf{F}}} =  \frac{\bra{ A_\mathsf{F}^\perp}  }{\sqrt{\brak{ A_\mathsf{F}^\perp}{ A_\mathsf{F}^\perp}}} \, \mathsf{S}  \,
 \frac{\ket{ A_\mathsf{F}} }{\sqrt{\brak{ A_\mathsf{F}}{ A_\mathsf{F}}}}
.
\EEA
Further, an explicit calculation gives:
\numparts
\BEA
X &= - i   \left[ \frac{\abs{a_1}^2 r_1^* \frac{\partial \, r_1}{\partial \theta}  + \abs{a_2}^2 r_2^* \frac{\partial \, r_2}{\partial \theta} }{ \abs{a_1 r_1}^2 + \abs{a_2 r_2}^2 }  -  \cot \Delta    \,\frac{ a_1 a_2 \left( r_1  \frac{\partial \, r_2}{\partial \theta} - r_2 \frac{\partial \, r_1}{\partial \theta} \right) }{ \abs{a_1 r_1}^2 + \abs{a_2 r_2}^2 } \right], \label{Res68X} \\
Y &=  i (r_1 + r_2) \cot \theta \left[\frac{a_2 a_1^* r_1^* - a_1  a_2^* r_2^*}{ \abs{a_1 r_1}^2 + \abs{a_2 r_2}^2 } + \cot \Delta \frac{a_1^2 r_1 + a_2^2 r_2}{ \abs{a_1 r_1}^2 + \abs{a_2 r_2}^2 }  \right] . \label{Res68Y}
\EEA
\endnumparts
We find that the enhancement term depends in general on the angle of incidence ($\theta$), on the polarization of the state ($a_1$ and $a_2$) as well as on the properties of the reflecting surface ($r_1$ and $r_2$).

\section{Conclusions}
In this work we have derived the spatial and angular Goos-H\"anchen and Imbert-Federov shifts of a beam with finite transversal extent after reflection using a quantum-mechanical notation. Studying these classical effects through the glasses of quantum mechanics gave some new insights. Our main result is equation \eref{Num7}. It furnishes that the spatial shift consist of two parts, one showing spatial-vs-angular shift mixing occurring, for example, for OAM beams. Furthermore, it becomes apparent that the angular shift is proportional to the beams angular spread (variance of the transverse component of the wave vector). Moreover, we studied the enhancement of beam shifts due to weak measurements and related our results to the seminal experiment of Hosten and Kwiat. The results presented here are in full agreement with the ones presented  by Dennis and G\"otte \cite{gotte1,gotte2}.

\end{document}